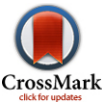

# Structure Evolution of Graphene Oxide during Thermally Driven Phase Transformation: Is the Oxygen Content Really Preserved?

**Pengzhan Sun[1], Yanlei Wang[2,3], He Liu[4], Kunlin Wang[1], Dehai Wu[4], Zhiping Xu[2,3], Hongwei Zhu[1,2]***

1 School of Materials Science and Engineering, State Key Laboratory of New Ceramics and Fine Processing, Key Laboratory of Materials Processing Technology of MOE, Tsinghua University, Beijing, P. R. China, 2 Center for Nano and Micro Mechanics, Tsinghua University, Beijing, P. R. China, 3 Department of Engineering Mechanics, Tsinghua University, Beijing, P. R. China, 4 Department of Mechanical Engineering, Tsinghua University, Beijing, P. R. China

## Abstract

A mild annealing procedure was recently proposed for the scalable enhancement of graphene oxide (GO) properties with the oxygen content preserved, which was demonstrated to be attributed to the thermally driven phase separation. In this work, the structure evolution of GO with mild annealing is closely investigated. It reveals that in addition to phase separation, the transformation of oxygen functionalities also occurs, which leads to the slight reduction of GO membranes and furthers the enhancement of GO properties. These results are further supported by the density functional theory based calculations. The results also show that the amount of chemically bonded oxygen atoms on graphene decreases gradually and we propose that the strongly physisorbed oxygen species constrained in the holes and vacancies on GO lattice might be responsible for the preserved oxygen content during the mild annealing procedure. The present experimental results and calculations indicate that both the diffusion and transformation of oxygen functional groups might play important roles in the scalable enhancement of GO properties.

**Data Availability:** The authors confirm that, for approved reasons, some access restrictions apply to the data underlying the findings. Relevant data are available at Figshare: http://dx.doi.org/10.6084/m9.figshare.1189083.

**Funding:** This work was supported by Beijing Natural Science Foundation (2122027), National Science Foundation of China (51372133), National Program on Key Basic Research Project (2011CB013000), Tsinghua University Initiative Scientific Research Program (2012Z02102) and the Training Program of Innovation and Entrepreneurship for Undergraduates (201410003B046). The funders had no role in study design, data collection and analysis, decision to publish, and preparation of the manuscript.

**Competing Interests:** The authors have declared that no competing interests exist.

* Email: hongweizhu@tsinghua.edu.cn

## Introduction

Graphene oxide (GO) [1–3], prepared by the oxidation and exfoliation of graphite, has been demonstrated to be an excellent 2D nanomaterial, which holds great promises for the next generation of nanodevices [4–6]. GO can be considered as graphene sheet asymmetrically decorated with oxygen-containing functional groups on the basal plane and the edges, resulting in a mixed $sp^2$-$sp^3$ carbon sheet [7,8]. Due to the chemical inhomogeneity and disordered structure of GO, it remains a challenge to understand the chemical structure and further to control the properties of GO [9–14]. So far, some significant studies have been conducted to understand the structure evolution of GO at various external stimuli such as temperature and harsh chemical environment. For example, Kim, et al. [15] have shown that GO is a metastable material and at room temperature, slow chemical and structural evolutions occur, resulting in a reduced O/C ratio and a structure deprived of epoxides and enriched in hydroxyls. Park, et al. [16] have demonstrated that hydrazine treatment of GO leads to the insertion of aromatic $N_2$ moieties in five-membered rings at the edges and the restoration of $sp^2$ graphitic regions on the basal planes. Bagri, et al. [17] have investigated the structure evolution of GO under progressive thermal reduction and demonstrated that high temperature annealing leads to the formation of carbonyl and ether groups through transformation of epoxides and hydroxyls, which hinders the complete removal of oxygen from GO. Unfortunately, the improvement of the sheet characteristics of GO through these typical chemical and thermal reduction routes comes at the expense of oxygen content. Considering the crucial importance of controlling the enriched and interactive oxygen networks on GO, which may give rise to the opening of band gaps for applications in electronics and photonics [8,12], a breakthrough has thus been made by Kumar, et al. [18], demonstrating that a direct mild annealing procedure can enhance the properties of GO on a large scale with the oxygen content preserved. Notably, they have proposed that a phase transformation process occurs by low-temperature driven oxygen diffusion on the basal plane that is responsible for the manipulation of the sheet properties of GO. However, the evolution of oxygen containing functional groups on GO during the mild annealing procedure is neglected in their work. Due to the metastable nature of GO [15], the transformation of diverse functional groups may occur during the low-temperature heating process, which might play an important role in the modulation of the properties of GO sheets.

In this work, the structure evolution of GO during the low-temperature annealing induced phase transformation is closely investigated by X-ray Photoelectron Spectrometer (XPS), Fourier Transform Infrared Spectroscopy (FTIR), X-ray Diffraction (XRD) and Auger Electron Spectroscopy (AES) analyses. The mechanism for the phase transformation process is further discussed based on the experimental results and density functional theory (DFT) based calculations, as illustrated in Fig. 1a. Finally, as an indirect evidence and a possible application, the modulation of ionic transport through GO membranes by thermally driven phase transformation is investigated.

## Results

GO sheets were synthesized by the modified Hummers' method [1–3]. As-prepared GO flakes were re-dispersed in water by sonication and drop-casted onto a piece of smooth paper followed by detached off to form the free-standing GO membranes (see Materials and Methods section) [19,20]. With these GO membranes, a mild low-temperature annealing (80°C) procedure was performed using a hot plate in air [18], during which the GO samples were taken out at regular intervals for structural analyses and subsequent ion permeation experiments. As shown in Fig. 1b, during the mild annealing process (80°C, 0~9 days), a gradual color change is evident in GO samples, which implies the existence of a clear structure evolution.

In order to study the structure evolution of GO during the low-temperature annealing process, C 1s XPS spectra were recorded with time and decomposed into five single Lorentzian peaks according to previous methods [21,22], which were assigned to C-C ($sp^2$), C-OH, C-O-C, C=O and O=C-OH respectively, as shown in Figs. 2a–c. It reveals that significant changes occur in the relative ratios of C-C ($sp^2$) and diverse oxygen functionalities. This clear structure evolution indicates that in addition to phase separation under temperature-driven oxygen diffusion on the basal plane [18], the transformation of oxygen containing functional groups also occurs, which might play an important role in the scalable enhancement of GO properties. The O 1s spectra of GO after heated at 80°C for 0 to 9 days are shown in Fig. 2d, indicating that during mild annealing, the O 1s peaks always locate at ~532.6 eV, corresponding to the contributions from C=O (531.2 eV), C(=O)-OH (531.2eV) and C-O (533 eV) [23,24]. The relative ratios of C and O elements were extracted from the XPS survey spectra, as plotted in Fig. 2e. It reveals that the O content of GO remains nearly unchanged during the mild heating process, except the slight decrease after 1-day annealing. This decrease in O content might be attributed to the loss of intercalated water within the membranes, which is in agreement with the previous work [18]. The loss of intercalated water within GO laminates can also be confirmed by XRD analysis, as shown in Fig. S1. It reveals that during the thermal annealing process, the diffraction peaks of GO membranes shift to higher angle values with time, corresponding to a gradual decrease in the interlayer spacing, which can be attributed to the loss of water in GO membranes. The relative ratios of C-C ($sp^2$) and various oxygen functional groups were further calculated with respect to the total area of the C 1s peak and the results are shown in Fig. 2f. Notably, it reveals that a significant change in the atomic percentages of oxygen functional groups occurs. In detail, during the low-temperature heating process, the ratios of $sp^2$ graphitic regions increase gradually, while C-OH and C=O decrease significantly. At the same time, the content of C-O (epoxy/ether) remains nearly unchanged, while the amount of carboxyl functional groups (COOH) has a slight increase. Based on the relative changes in C-C ($sp^2$) and various oxygen functionalities (Fig. 2f), it can be concluded that during this thermally treated procedure, the amount of O chemically bonded to C atoms decreases continuously, which is in contradiction to the conclusions drawn by Kumar, et al. [18] Considering the fact that the relative contents of C and O are nearly unchanged (>1 day annealing, Fig. 2e), it can be inferred that the strongly physisorbed O species that are constrained in the carbon vacancies and holes in GO lattice contribute to the nearly unchanged O content [24–28].

The structure changes of GO during the thermal annealing process were also closely examined by FTIR. Considering the nearly constant C-O (epoxy/ether) ratios shown in Fig. 2f, the FTIR spectra were all normalized by the intensity of C-O located at ~1225 $cm^{-1}$ and the results are shown in Fig. 3a. It reveals that the intensities of the bands assigned to C-OH (~3400 $cm^{-1}$) and C=O (carboxyl/carbonyl, ~1730 $cm^{-1}$) weaken gradually with annealing (Fig. 3b), which is in consistent with the XPS data in Fig. 2f. As the relative content of carboxyl is much smaller than that of carbonyl on GO (Fig. 2f), the change tendency of the band intensity assigned to C=O is mainly dominated by carbonyl, further leading to the gradual decrease in the intensity of the C=O band located around 1730 $cm^{-1}$, as shown in Fig. 3. In addition, it is seen that the peak located at ~1625 $cm^{-1}$, which is assigned to the C-C skeletal vibrations or the deformation vibration of intercalated water, gradually decreases, indicating the loss of intercalated water [18]. Instead, a new peak located below 1600 $cm^{-1}$ appears with increasing intensity, which is attributed to the formation of graphitic domains [18]. These results are further in consistent with the XRD analysis in Fig. S1 and the XPS data in Fig. 2.

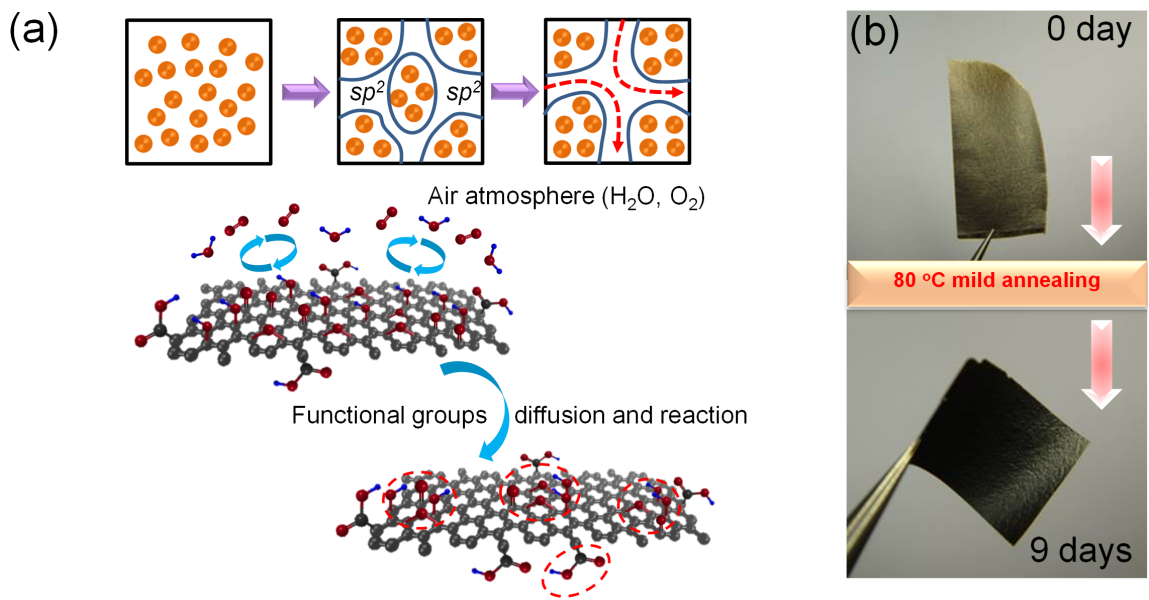

**Figure 1. GO membranes.** (a) Schematic diagram for the diffusion and transformation of oxygen functionalities on GO during the mild annealing procedure. (b) Photographs of GO samples after annealing at 80°C for 0 and 9 days, respectively.
doi:10.1371/journal.pone.0111908.g001

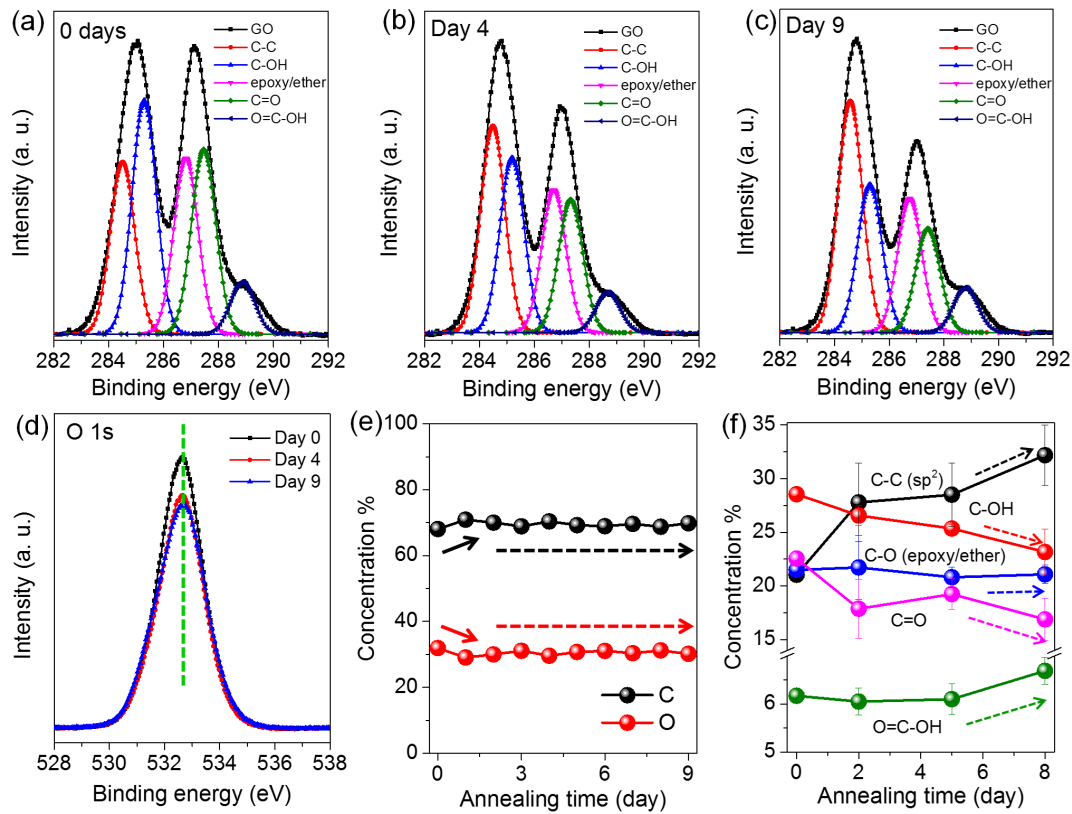

**Figure 2. XPS spectra of GO during the mild annealing procedure at 80°C for (a) 0 day, (b) 4 days and (c) 9 days, respectively. (d) The O 1s spectra of GO after annealing for 0 to 9 days. (e) The atomic percentages of C and O after annealing for 0 to 9 days. (f) The relative ratios of C-C ($sp^2$) and diverse oxygen functional groups during the mild annealing process.**
doi:10.1371/journal.pone.0111908.g002

## Discussion

Bagri, *et al.* [17] have reported the structure evolution of GO under progressive thermal reduction, demonstrating that high temperature treatment leads to the formation of carbonyl and ether functional groups. In the present work, the XPS and FTIR results co-demonstrate the decreased percentage of C=O and the constant amount of C-O groups under low-temperature annealing, which is in contradiction to the high temperature case. On the other hand, Kim, *et al.* [15] have demonstrated that GO is metastable and it can undergo a structural and chemical evolution at room-temperature to reach a quasi-equilibrium, where GO reaches a reduced O/C ratio and a structure deprived of epoxides and enriched in hydroxyls. These conclusions are further opposite to the case here, where we show that during the mild thermal treatment, the amount of C-OH decreases while the amount of C-O remains nearly constant (Figs. 2,3). Therefore, a new mechanism responsible for the structure evolution of GO under mild annealing should be proposed.

To explore the possible migration and transition paths of oxygen-rich functional in graphene oxide, we performed DFT based first-principles calculations. Elementary binary reactions among hydroxyl, epoxide and carbonyl species chemisorbed on graphene were considered, following a recent work reported by Zhou *et al.* [29] The GO structure is modeled as a 5×6 super-cell of graphene, functionalized by specific oxygen-rich groups. The atomic structures, their energy differences and reaction barriers were obtained here by DFT based calculations, as summarized in Fig. 4. The plane-wave code Quantum-Espresso was used here with an energy cutoff of 70 Ry. Norm-conserving pseudopotentials [30] was used for the core-valence electron interaction and the Perdew-Burke-Ernzerhof (PBE) parametrization of generalized gradient approximation (GGA) was implemented for the exchange-correlation functional [31]. These settings have been verified to achieve a total energy convergence for the systems under exploration below 1 meV/atom. The energy difference of a reaction $\Delta E = E_{prod} - E_{react}$ was calculated from relaxed structures of the reactant and product, respectively. The activation reaction barrier $E_b$ was then calculated using the nudged elastic band (NEB) technique. A $10\times10\times1$ Monkhorst-Pack mesh grid was set for $k$-point sampling, while only $\Gamma$-point sampling was used for the reaction path searching, to save the computational demand. The vacuum region in our super-cell approach was set to 1.2 nm in the direction perpendicular to the basal plane of graphene. The energy diagrams in Fig. 5 show that the reaction from the carbonyl and hydrogen pairs to hydroxyl pairs is exothermic with $\Delta E = -2.45$ eV and $E_b = 2.33$ eV, while the reaction from a carbonyl pair to the epoxides is endothermic with $\Delta E = 1.07$ eV and $E_b = 1.90$ eV. Notably, the formation of a carboxyl group from carbonyl and hydroxyl groups is exothermic with negligible $\Delta E = -0.013$ eV and $E_b = 0.31$ eV.

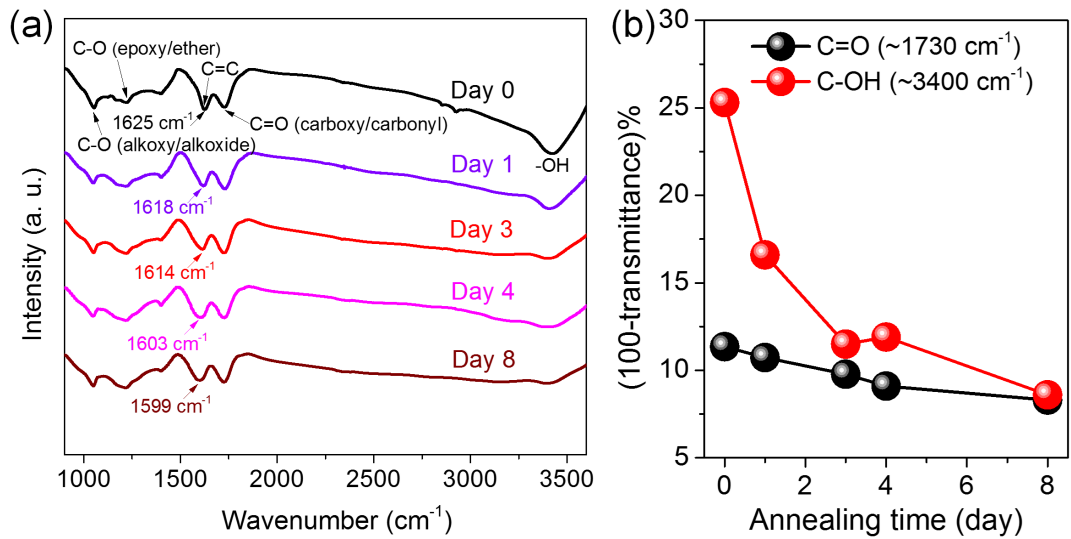

**Figure 3. FTIR spectra of GO membranes after low-temperature annealing for 0 to 8 days, showing the relative changes of oxygen functionalities.** The plots in (b) show the transmittance changes of the bands assigned to C=O (~1730 cm$^{-1}$) and C-OH (~3400 cm$^{-1}$) during mild annealing.
doi:10.1371/journal.pone.0111908.g003

It has been demonstrated experimentally and theoretically that during the mild annealing process, the interactions between epoxides and hydroxyls are attractive [29] and they can undergo a significant diffusion process, resulting in the gradual separation of $sp^2$ and $sp^3$ phases along the graphene basal plane [18]. The diffusion rates of isolated epoxides and hydroxyls are controlled by activation barriers of 0.8 eV and 0.3 eV, respectively [15,18,29]. At 80°C, the diffusions of hydroxyls and epoxides are also expected to increase by 1 and 2 orders of magnitude, respectively, compared to the room temperature case [18], which facilitates the rapid migration and aggregation of epoxides and hydroxyls along the surfaces. Such aggregation of oxygen functionalities under mild annealing should lead to the gradual increase of $sp^2$ graphitic domain sizes, as demonstrated by the XPS and FTIR results in Figs. 2 and 3. During the phase separation process, the transformations among diverse functional groups are also expected to occur, as shown in Figs. 2 and 3. Particularly, several reaction strategies which are presumably responsible for the structure evolutions involved in the mild annealing procedure are proposed as follows: (i) The intercalated water molecules that are constrained in the holes or vacancies on GO can be dissociated into C-OH, C=O and C-H groups through interacting with the active edge carbon atoms with an energy barrier of ~0.7 eV [26]. Then C-OH and C-H species tend to diffuse along the basal plane with a barrier of ~0.3 eV and ~0.5 eV, respectively [15]. (ii) Due to the higher energy states of carbonyl pairs with chemisorbed H species nearby [29], they can evolve into hydroxyl pairs (2.33 eV relative to a carbonyl pair, reaction a in Fig. 4) or an epoxide and a hydroxyl with the assistance of C-H groups [29]. They may also directly transform into epoxide pairs with an energy barrier of 1.9 eV (relative to a carbonyl pair, reaction b in Fig. 4). Although this reaction is endothermic, the energy could be lowered as the epoxides further diffuse away with an activation barrier of 0.8 eV. These possible reactions presumably lead to the gradual decrease of the amount of C=O groups (Figs. 2 and 3). (iii) In the presence of C-H, hydroxyls and epoxides can react readily with the H species to produce H$_2$O molecules with an energy barrier $\leq$0.15 eV [15]. Two hydroxyls can also attract each other and interact to produce an epoxide and release a H$_2$O molecule by overcoming a barrier of 0.5 eV [29]. These reactions might give rise to the decreased ratio of C-OH groups (Figs. 2 and 3). On the other hand, two epoxides on the same side of graphene can react with each other to produce an O$_2$ molecule with an energy barrier of 1.0 eV while an epoxide pair on opposite sides of graphene can also interact and lead to the formation of a carbonyl pair with a barrier of 0.8 eV [29]. The reversible interactions of epoxides with carbonyl and hydroxyl functional groups are presumably responsible for the nearly constant ratio of epoxide groups, as demonstrated in Fig. 2. (iv) Notably, our simulation results also demonstrate that a hydroxyl and a carbonyl group at the sheet edge can react readily with each other to form a carboxyl group with an energy barrier of 0.31 eV (reaction c in Fig. 4), further supporting our experimental results that during the mild annealing process, the amount of carboxyl groups decorated on the sheet edges increases slightly, which can also contribute to the decrease of hydroxyl and carbonyl groups, as shown in Figs. 2 and 3. Moreover, the slight increase of carboxyl at the edges is also a direct evidence for the diffusion of oxygen functional groups along the graphene basal plane.

Next, in order to directly probe the phase transformation processes in GO membranes, AES characterizations on GO samples after annealing at 80°C for 0 and 9 days respectively were performed, as shown in Fig. 5. It reveals in Figs. 5a–c that during the mild annealing process, the O content within GO membranes indeed decreases gradually, while the concentration of O species physically adsorbed on GO surfaces increases with annealing.

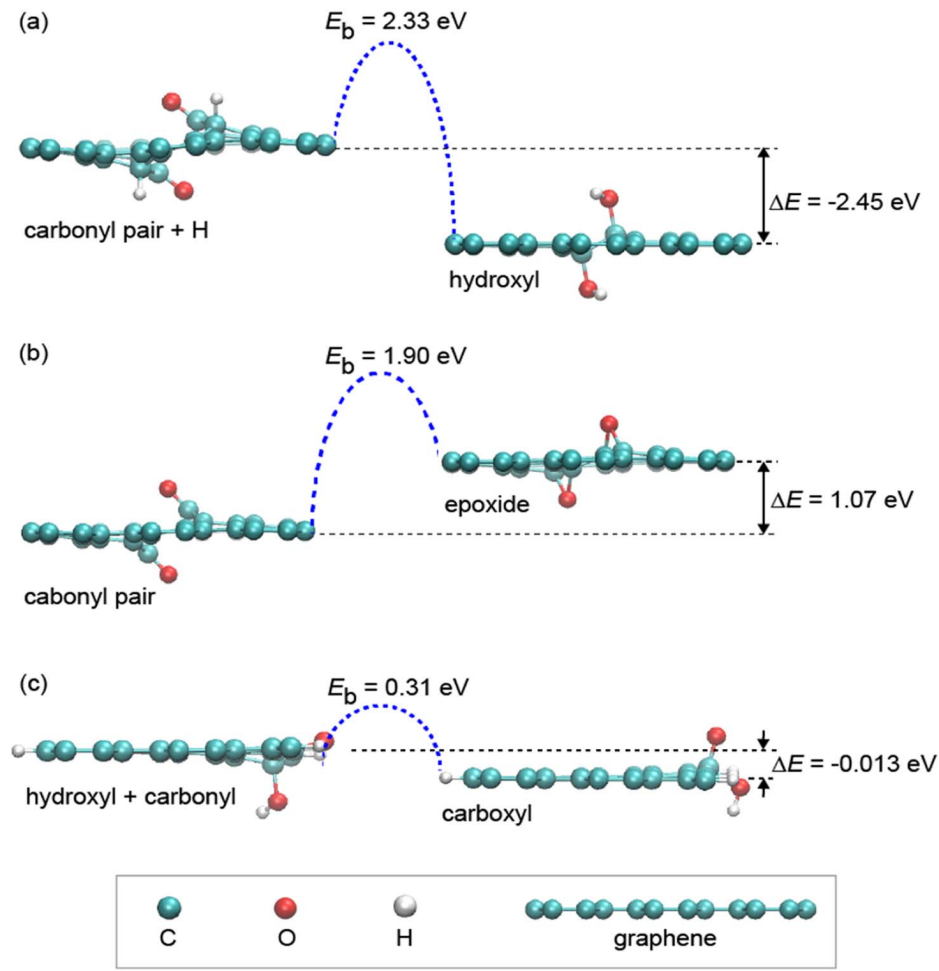

**Figure 4. Formation energies and reaction barriers for proposed evolution paths (a–c) of oxygen containing functional groups on GO, calculated by DFT based calculations.**
doi:10.1371/journal.pone.0111908.g004

These results further demonstrate our conclusion drawn from the XPS analyses in Fig. 2 that during the mild annealing process, the amount of O atoms chemically bonded to the carbon basal plane decreases gradually and the strongly physisorbed external O species in GO lattice contribute to the nearly unchanged O content. Figs. 5d and e exhibit the C and O mappings (20 μm×20 μm) of GO membranes after annealing for 0 and 9 days, respectively. In order to exclude the effect of physisorbed O species on the surfaces, the C and O mappings were performed after sputtering the GO membranes for 5 min (Sputter rate: 35 nm/min). Clearly, it reveals in Fig. 5d that the as-prepared GO membrane shows a relative uniform C and O distribution. The graphene clusters are small and they distribute randomly on the GO basal plane. In contrast, after mild annealing for 9 days, it is seen from Fig. 5e that the C-C regions gradually aggregate and coalesce with each other to result in larger graphene-rich domain sizes, further demonstrating the occurrence of phase separation process during the low-temperature annealing procedure. Notably, by comparing the C and O maps for GO membranes after annealing for 0 and 9 days (Figs. 5d and e), we can also conclude that the amount of O chemisorbed on the carbon lattice gradually decreases with annealing, which is again in consistent with the XPS (Fig. 2f), FTIR (Fig. 3) and AES (Figs. 5a–c) results, but in contradiction to the work by Kumar *et al*. [18], where the strong physisorption was not excluded.

Finally, as an indirect evidence and a possible application for the structure evolution of GO, the modulation of ion permeations through GO membranes after mild annealing at 80°C for 0∼9 days was investigated based on MgCl$_2$ sources, which represent a typical kind of chloride solutions [32] and avoid the coordination interactions between metal cations and oxygen functionalities within the $sp^3$ C-O matrix [19,20]. Inspired by the work of Kumar *et al*. [18], it was believed that the diffusion and aggregation of oxygen functionalities on GO basal planes should increase the sizes of $sp^2$ domains (illustrated in Fig. 1a), which would further increase the amount of continuous and smooth $sp^2$ nanocapillaries within the GO membranes and facilitate the faster transport of various ions. Herein, the ion permeation experiments were done with a home-made apparatus, as illustrated in Figs. 6a and b (see Materials and Methods section for detailed information). The ionic transportations through GO membranes after mild annealing for various degrees are shown in Fig. 6c. Anomalously, it is seen that the permeances of source ions decrease gradually with annealing, which is just in contradiction to the inference drawn from the previous work by Kumar, *et al*. [18] The permeation rates of Mg$^{2+}$ cations through GO membranes with various extents of annealing were further calculated based on the atomic emission spectroscopy data and the results are shown in Fig. 6d. Again it reveals that the ionic permeation rates decrease with the thermally driven phase transformation process, which is also opposite to the initial predication. These results indicate that in addition to oxygen diffusion, the transformation of oxygen functionalities decorated on GO surfaces also occurs during the mild heating procedure, which should lead to the slight reduction and further the decrease of ion permeation rates through GO membranes. In detail, based on the above experimental results and DFT calculations, it can be concluded that during the mild annealing procedure, the thermally driven diffusion of oxygen species should increase the sizes of $sp^2$ graphitic domains (Fig. 2f and Fig. 5e), which would further lead to the increase of the amount of continuous and smooth $sp^2$ nanocapillaries across all the stacking layers within GO membranes (Fig. 1a). On the other hand, the transformation among diverse oxygen functionalities on GO surfaces should decrease the amount of chemically bonded oxygen atoms gradually, as demonstrated by the XPS analyses shown in Fig. 2 and the AES characterizations shown in Fig. 5. This slight reduction of GO membranes would lead to the decrease of the interlayer spacing between two GO layers (Fig. S1) and further give rise to the decrease of the ionic transport through GO membranes [33]. In addition, the decrease of ion permeation through GO membranes might also originate from the effective capture of water molecules by the larger oxidized regions formed during the thermally driven phase separation process under mild annealing [34], which results in the slower permeation of ions and water molecules through GO membranes.

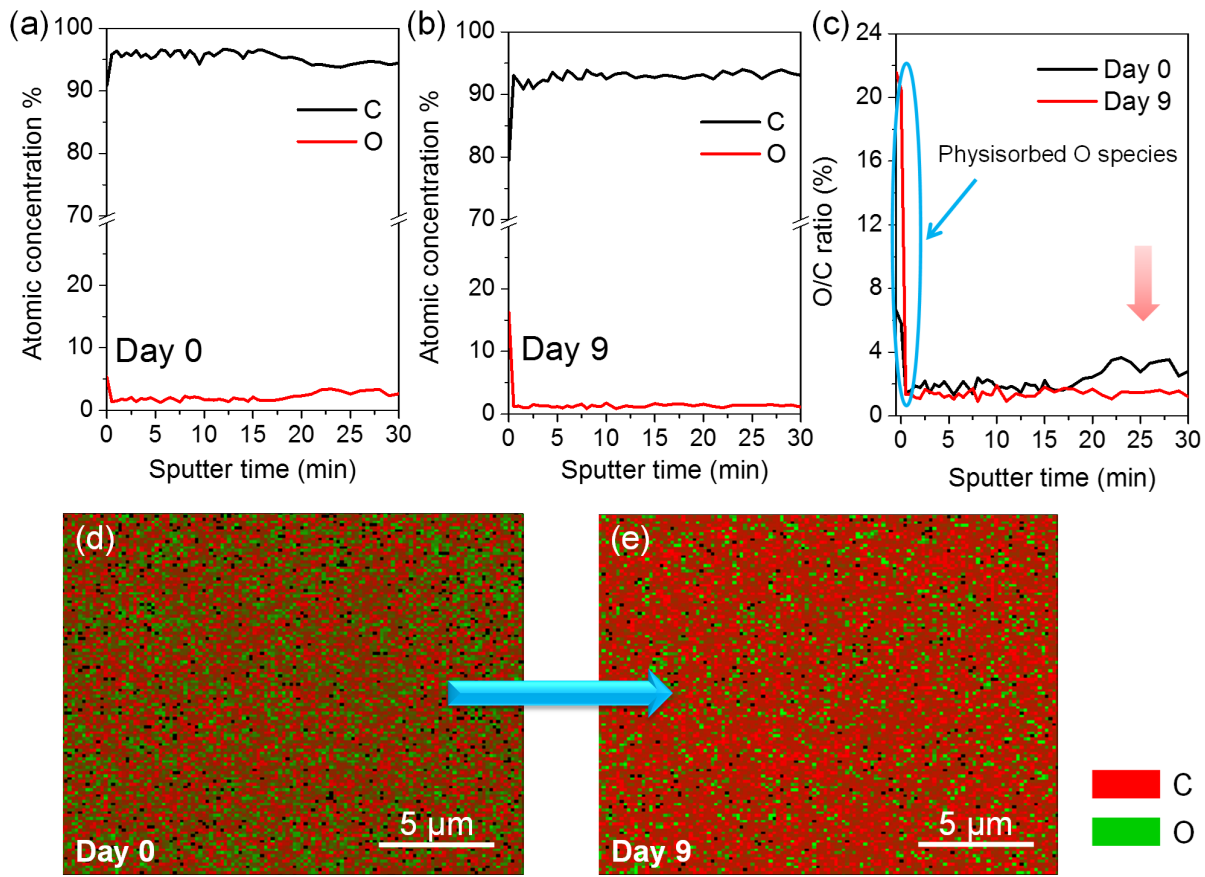

**Figure 5. AES characterizations (the C and O concentration distributions along the thickness) for GO membranes after annealing at 80°C for (a) 0 and (b) 9 days, respectively.** The sputter rate is 35 nm/min. (c) The comparison of O/C ratio distributions along the depth within GO membranes after 0- and 9-day annealing. (d, e) C (red) and O (green) mappings of GO membranes after 0- and 9-day annealing, respectively. The mappings were performed after sputtering the GO membranes for 5 min to exclude the effect of surface O physisorption.
doi:10.1371/journal.pone.0111908.g005

## Conclusions

In summary, the structure evolution of GO under mild annealing is closely investigated *via* XPS, FTIR, XRD and AES analyses. The results indicate that in addition to phase separation, significant transformation among diverse oxygen functionalities also occurs, which leads to the slight reduction of GO membranes and further the enhancement of GO properties. These results are further supported by the NEB-DFT calculations. Notably, the amount of O chemically bonded to C atoms decreases gradually and we propose that the intercalated O species that are constrained in the small holes and vacancies on GO lattice are responsible for the preserved O content as reported by Kumar, *et al.* [18] The results present here provide insight into the fundamental mechanism that involves in the mild annealing of GO and further indicate that both the diffusion and transformation of oxygen species might play important roles in the scalable enhancement of GO properties during the low-temperature thermally treated procedure.

## Materials and Methods

### Free-standing GO membrane preparation

GO flakes were synthesized by the typical modified Hummers' method using sodium nitrite, potassium permanganate and concentrated sulfuric acid according to previous work [1–3]. The as-prepared GO sheets were re-dispersed in water by sonication to form the 1.5 mg/mL aqueous solutions. After that, GO preparation droplets (∼1 mL, 1.5 mg/mL) were drop-casted onto a piece of smooth paper, followed by drying spontaneously and detaching off to form the free-standing GO membranes [19,20], which were utilized for structural analyses and ionic trans-membrane permeation experiments subsequently.

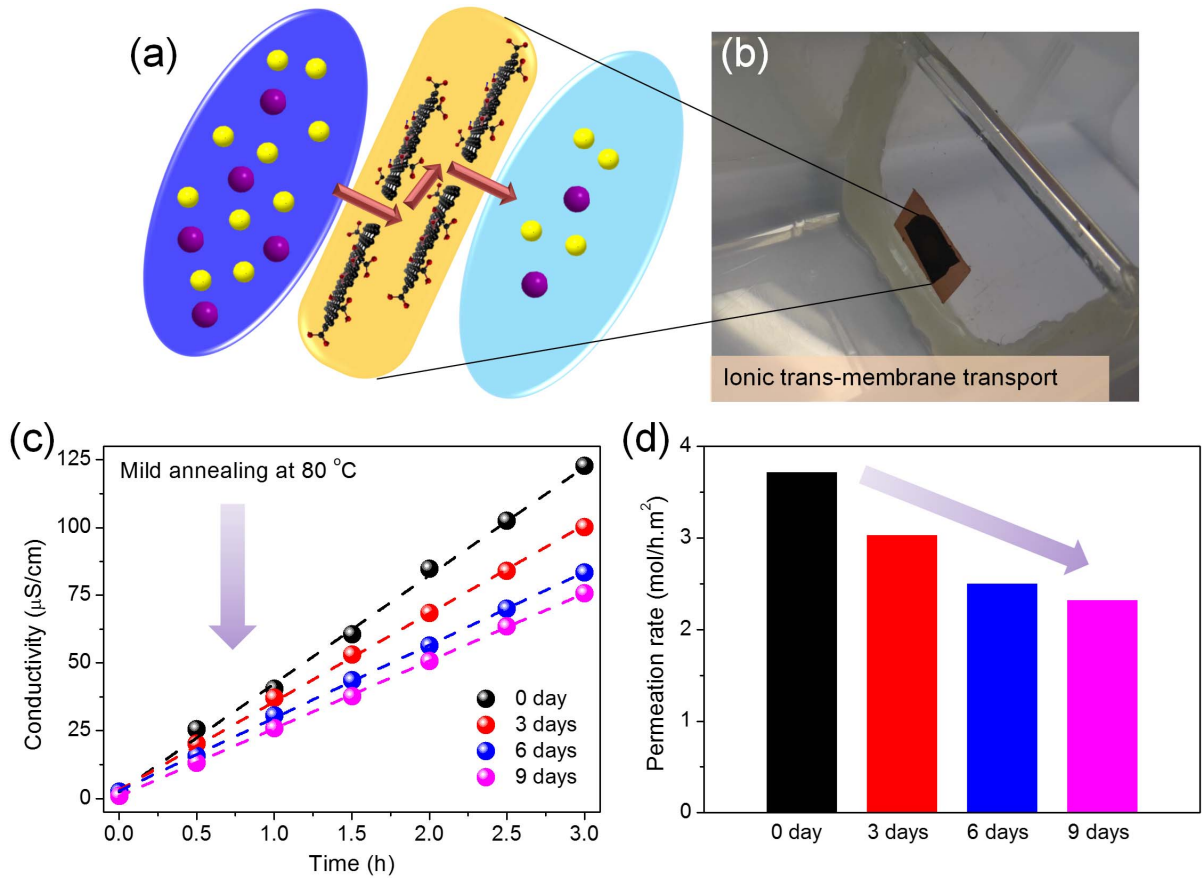

**Figure 6. Ion permeation tests.** (a) Schematic drawing for the ionic transport through GO laminates. (b) Photograph of the self-made ion permeation apparatus. (c) Ionic permeation processes (drain conductivity variations *versus* time) through GO membranes after mild annealing for 0 to 9 days. (d) The changes of ion permeation rates for GO membranes annealing for various degrees.
doi:10.1371/journal.pone.0111908.g006

### Ionic permeation experiments

The ionic permeation experiments were conducted with a self-made penetration apparatus according to previous method [19,20]. Briefly, a piece of GO membrane was sealed with double-sided copper tape onto an aperture (5 mm in diameter) on the plate which separated the source vessel from the drain vessel. This facilitated the direct connection of source and drain solutions by GO membranes and the trans-membrane transportation of source ions without passing through any supporting substrates. During the penetration experiments, 80 mL, 0.1 mol/L $MgCl_2$ solutions and deionized water were injected into the source and drain vessels respectively with the same speed and the conductivity variations of the drains were measured on a conductivity meter (INESA, DDS-307) with time under mild stirring to reflect the trans-membrane permeation behaviors of source ions. After permeating the GO membranes for 3 h, the filtrates were collected for atomic emission spectroscopy analyses (IRIS Intrepid II) for accurate concentrations of $Mg^{2+}$, from which the permeation rates (the amount of ions transported per hour per unit area) of source cations could be calculated.

### Structural analyses

The structure evolution of GO membranes during the mild annealing procedure was monitored by X-ray photoelectron spectroscopy (XPS, PHI Quantera SXM, AlKα), Fourier Transform Infrared Spectroscopy (FTIR, Nicolet 6700FTIR), X-ray diffraction (XRD, Siemens, 08DISCOVER, λ = 0.15405 nm) and Auger Electron Spectroscopy (AES, PHI-700) techniques.

### Supporting Information

**Figure S1 XRD diffractograms of GO laminates during the low-temperature annealing process showing the changes of interlayer distances.**
(TIF)

### Author Contributions